\documentclass[journal,twoside,web]{ieeecolor}

\usepackage{tmi}
\usepackage{cite}
\usepackage{amsmath,amssymb,amsfonts}
\usepackage{graphicx}
\usepackage{textcomp}
\usepackage{hyperref}
\usepackage{listings}
\usepackage{booktabs, multirow}
\usepackage{siunitx}
\usepackage{algorithm}
\usepackage{algpseudocode}

\begin{document}

\title{Benchmarking of Deep Learning Methods \\for Generic MRI Multi-Organ \\Abdominal Segmentation}

\author{Deepa Krishnaswamy, Cosmin Ciausu, Steve Pieper, Ron Kikinis, Benjamin Billot, Andrey Fedorov
\thanks{This work was supported by the NIH NCI (HHSN26110071, HHSN261201500003l) and used Jetstream2 GPU at Indiana University, which is supported by National Science Foundation grants (2138259, 2138286, 2138307, 2137603, 2138296).}
\thanks{D. Krishnaswamy and C. Ciausu contributed equally to this work.}
\thanks{D. Krishnaswamy (dkrishnaswamy@bwh.harvard.edu, corresponding author) is with Brigham and Women's Hospital, Boston, USA.}
\thanks{S. Pieper (pieper@isomics.com)is with Isomics, Cambridge, USA.}
\thanks{C. Ciausu (cciausu@bwh.harvard.edu), R. Kikinis (kikinis@bwh.harvard.edu), and A. Fedorov (afedorov@bwh.harvard.edu) are with Brigham and Women's Hospital, Boston, USA.}
\thanks{B. Billot is with Inria, Epione team, Sophia-Antipolis, France (benjamin.billot@inria.fr).}}
\maketitle

\begin{abstract}  % 250 words or less.
Recent advances in deep learning have led to robust automated tools for segmentation of abdominal computed tomography (CT). Meanwhile, segmentation of magnetic resonance imaging (MRI) is substantially more challenging due to the inherent signal variability and the increased effort required for annotating training datasets. Hence, existing approaches are trained on limited sets of MRI sequences, which might limit their generalizability. To characterize the landscape of MRI abdominal segmentation tools, we present here a comprehensive benchmarking of the three state-of-the-art and open-source models: \textit{MRSegmentator}, \textit{\mbox{MRISegmentator-Abdomen}}, and \textit{TotalSegmentator MRI}. Since these models are trained using labor-intensive manual annotation cycles, we also introduce and evaluate \textit{ABDSynth}, a SynthSeg-based model purely trained on widely available CT segmentations (no real images). More generally, we assess accuracy and generalizability by leveraging three public datasets (not seen by any of the evaluated methods during their training), which span all major manufacturers, five MRI sequences, as well as a variety of subject conditions, voxel resolutions, and fields-of-view. Our results reveal that \textit{MRSegmentator} achieves the best performance and is most generalizable. In contrast, \textit{ABDSynth} yields slightly less accurate results, but its relaxed requirements in training data make it an alternative when the annotation budget is limited. The evaluation code and datasets are given for future benchmarking at \url{https://github.com/deepakri201/AbdoBench}, along with inference code and weights for \textit{ABDSynth}.
\end{abstract}


\begin{table*}[t]
\caption{Summary of the training data for the benchmarked methods. Brackets denote ranges. CE=contrast-enhanced.}
\centering
\fontsize{7.5}{5}\selectfont
\setlength\tabcolsep{2pt}
\begin{tabular}{lccccccc}
\toprule
Method & Training source & \# scans & Abnormalities  & Data type & Resolution (mm) & Dimension \\
\midrule
\multirow{9}{*}{MRSegmentator \cite{Häntze2024}}  
    & \multirow{4}{*}{UK Biobank \cite{Sudlow2015}} 
    & \multirow{4}{*}{1200} 
    & \multirow{4}{*}{various pathologies}
          & T1 Dixon in-phase       &                   &  \\
    & & & & T1 Dixon out-of-phase   & [2.23×2.23×3.00]  & [224×156×44] \\
    & & & & T1 Dixon water only     &  [2.23×2.23×4.50] &  [224×174×72] \\
    & & & & T1 Dixon fat only       &                   & \\
\cmidrule{4-7}
    & \multirow{3}{*}{In-house dataset} 
    & \multirow{3}{*}{221} 
    & \multirow{3}{*}{kidney tumors} 
          & T1                 & \multirow{3}{*}{1.00×1.00×1.00} & 100–450 \\
    & & & & T1 fat-saturated   & & (only given in \\
    & & & & T2 fat-saturated   & & axial direction)\\
\cmidrule{4-7}
    & \multirow{2}{*}{TotalSegmentator \cite{Wasserthal2023, WasserthalZenodo}} 
    & \multirow{2}{*}{1228} 
    & \multirow{2}{*}{various pathologies} 
    & \multirow{2}{*}{CT}                 
    & \multirow{2}{*}{1.50×1.50×1.50} 
    & [47×48×29] \\
    & & & & & & [499×430×851] \\
\midrule
\multirow{4}{*}{MRISegmentator-Abdomen \cite{Zhuang2024}}  
    & \multirow{4}{*}{In-house dataset} 
    & \multirow{4}{*}{780} 
    & & T1 pre-contrast       & [0.94×0.94] &  \\
    & & & liver tumors, pancreatic cysts & CE T1 arterial phase & [1.47×1.47] & [228×240×80] \\
    & & & other abnormalities & CE venous phase      &  (inter-slice res.             & [320×320×96]\\
    & & & & CE delayed phase     &   not given)            & \\
\midrule 
    & \multirow{3}{*}{University Hospital Basel} 
    & \multirow{9}{*}{} 
    & \multirow{3}{*}{various pathologies}
          & T1               & [0.17×0.17×0.17] & [11×10×10] \\
    & & & & T2               & [20.0×10.64×14.40] & [1092×1280×1915] \\
    & & & & Proton density   &                   & \\
\cmidrule{4-7}
\multirow{2}{*}{TotalSegmentator MRI \cite{D'Antonoli2024}}    & \multirow{2}{*}{Imaging Data Commons \cite{Fedorov2023}} 
    & \multirow{2}{*}{1088} 
    & \multirow{2}{*}{cancer}
    & \multirow{2}{*}{Various sequences} & [0.29×0.29×0.43] & [17×5×5] \\
    & & & &                              & [7.50×25.00×28.0]  & [672×672×512] \\
\cmidrule{4-7}
    & \multirow{2}{*}{TotalSegmentator \cite{Wasserthal2023, WasserthalZenodo}} 
    & \multirow{2}{*}{} 
    & \multirow{2}{*}{various pathologies}
        & \multirow{2}{*}{CT}               & \multirow{2}{*}{1.50x1.50×1.50} & [47×48×29] \\
    & & & &                                   &                              & [499×430×851] \\
\midrule
ABDSynth 
    & Subset of TotalSegmentator
    & 128 
    & various pathologies
    & CT segm. (no real images) & 1.50×1.50×1.50 & 300×300×250 \\
\bottomrule
\end{tabular}
\label{tab:eval_methods}
\end{table*}

%%%%%%%%%%%%%%%%%%%%%%%%%%%%%%%
\section{Introduction}
%%%%%%%%%%%%%%%%%%%%%%%%%%%%%%%

\IEEEPARstart{A}{ccurate} segmentation of abdominal organs in magnetic resonance imaging (MRI) volumes is a prerequisite for an array of clinical tasks \cite{Lenchik2019} such as volumetry \cite{DeSouza2018, Zöllner2012}, early diagnosis, longitudinal disease monitoring, radiotherapy planning \cite{Dirix2014, Keall2022}, and biomarker extraction \cite{Gillies2016, Barat2024}. However, manual expert contouring is labor-intensive \cite{Heerkens2017} and prone to inter- and intra-rater reproducibility issues \cite{Podobnik2024}. To mitigate these challenges and improve consistency, automated multi-organ segmentation methods have been proposed. 

While automated segmentation of abdominal MRI scans is essential, this task has been hindered by the large variability in acquisition parameters for this modality. Indeed, MRI lacks inherent intensity normalization, which poses a challenge in automatic segmentation since traditional deep neural networks \cite{Ronneberger2015} are fragile against MRI contrast variations \cite{billot_2020_learning}, an issue known as ``domain gap'' \cite{pan_2010_survey}. This variability also complicates manual MRI segmentation and thus hinders the creation of large annotated training sets. As a result, only a few methods have been proposed until recently for multi-organ abdominal MRI segmentation \cite{Chen2020, Kart2021, Rickmann2022, Amjad2023}. This is in contrast with computed tomography (CT), where methods like TotalSegmentator can benefit from the highly standardized signal \cite{Wasserthal2023}.

Recent advances in deep learning segmentation networks, best represented by nnU-Net \cite{Isensee2021}, have led to the development of state-of-the-art methods for multi-organ abdominal segmentation in MRI: \textit{MRSegmentator} \cite{Häntze2024}, \textit{\mbox{MRISegmentator-Abdomen}} \cite{Zhuang2024}, and \textit{TotalSegmentator MRI} \cite{D'Antonoli2024}, each capable of segmenting more than 40 regions, including organs, bones, muscles, and vessels. The development of these methods has been performed jointly with the annotation of their respective training datasets. Specifically, the employed training procedures all rely on a very labor-intensive strategy where annotated datasets are obtained with iterative expert refinements. With this technique, the aforementioned methods are trained on large MRI cohorts from the UK Biobank \cite{Sudlow2015}, the German National Cohort (NAKO) \cite{NAKO2014}, Imaging Data Commons \cite{Fedorov2023}, TotalSegmentator \cite{Wasserthal2023}, and other sources \cite{Häntze2024, Mathai2022, Mathai2023}. Crucially, these datasets span multiple MRI sequences, manufacturers, voxel resolutions, and pathologies, thereby improving the robustness of these methods compared to previous approaches \cite{Chen2020,Kart2021} by increasing the variability of the training data. However, the accuracy and generalizability of these models remain to be compared for generic out-of-the-box usage.

Since direct annotation of MRI datasets is challenging, other approaches propose to tackle the generalization issue of abdominal MRI segmentation networks by adopting domain adaptation strategies \cite{kondrateva2021domain}, where the goal is to transfer models trained on labeled data from a source domain to a target domain where labels are unavailable. Li et al. propose training an image-translation network to create synthetic MRI data with pseudo-labels \cite{Li2024}. Segmentation is then performed with an automatic multi-stage network. Another approach \cite{Wu2024} first performs CT-to-MRI translation using a CycleGAN model \cite{CycleGAN2017} equipped with an organ-attention mechanism. An nnU-Net with some additional enhancements is then trained for supervised segmentation of these MRI scans. However, these domain adaptation approaches need to be retrained for each new MRI sequence, which does not comply with our scenario of out-of-the-box MRI segmentation. 

SynthSeg \cite{Billot2023,billot_2023_robust} is a relatively new technique that proposes to circumvent domain adaptation strategies with domain randomization \cite{Tobin2017}. Specifically, SynthSeg leverages a parametric generative model based on a Gaussian mixture model (GMM) conditioned on input segmentations (no real images needed). Synthetic scans are created by randomly sampling all generation parameters from wide uniform distributions, thus yielding scans of randomized MRI contrast. Exposing a downstream segmentation network to such variable scans forces it to learn domain-agnostic features, so that the trained network can be used on any domain without retraining \cite{Billot2023}. SynthSeg has originally been proposed for brain segmentation \cite{billot_2020_learning}, but has since been extended to MRI and CT cardiac segmentation \cite{Billot2023}. More generally, the SynthSeg framework is an alternative to the latest methods in MRI abdominal segmentation, since it eases their burdensome annotation process by only requiring segmentations as training inputs, which can be taken from other modalities such as widely available CT label maps \cite{Wasserthal2023}.

In this paper, we propose a benchmark for existing and future abdominal MRI segmentation methods. In particular, we conduct an analysis on three datasets: AMOS MRI \cite{Ji2022}, CHAOS MRI \cite{Kavur2019,Kavur2020,Kavur2021}, and LiverHCCSeg \cite{Gross2023,LiverHccSeg}. These contain a variety of MRI scans acquired at different institutions, using scanners from all major manufacturers, and include variable MRI sequences, voxel resolutions, and populations (i.e., healthy subjects and diseased patients). Here, we assess the performances of the three state-of-the-art methods: \textit{MRSegmentator} \cite{Häntze2024}, \textit{\mbox{MRISegmentator-Abdomen}} \cite{Zhuang2024}, and \textit{TotalSegmentator MRI} \cite{D'Antonoli2024}. For completeness, we also evaluate a new method, named \textit{ABDSynth}, that extends SynthSeg \cite{Billot2023} for out-of-the-box MRI multi-organ abdominal segmentation and that is trained solely on CT segmentations. Our results reveal that \textit{MRSegmentator} outperforms the other methods, both for in-domain accuracy and out-of-domain generalization. Meanwhile, \textit{ABDSynth} is slightly less accurate than the other methods, but presents an alternative in scenarios where annotated MRI data is scarce. The data and evaluation code are available for future benchmarking  \url{https://github.com/deepakri201/AbdoBench}.

%%%%%%%%%%%%%%%%%%%%%%%%%%%%%%%
\section{Materials and methods}
%%%%%%%%%%%%%%%%%%%%%%%%%%%%%%%

\begin{figure*}[t]
\centering
\includegraphics[width=0.9\textwidth]{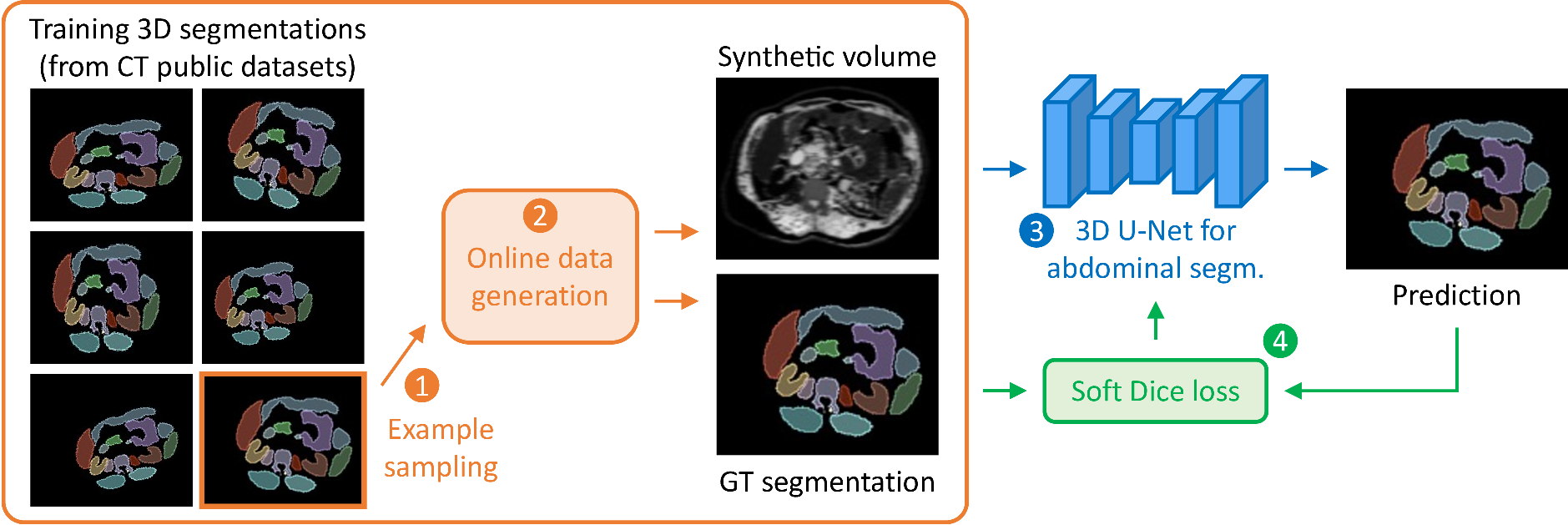} 
\caption{Overview of an \textit{ABDSynth} training step. 1) A CT segmentation is sampled from the training set. 2) A synthetic volume is generated using a segmentation-conditioned GMM with randomized parameters. 3,4) Abdominal volume/segmentation pairs are used to train a supervised 3D U-Net.}
\label{fig:overview}
\end{figure*}

\begin{table*}[t]
\caption{Summary of the publicly available datasets used for evaluation. Brackets denote ranges.}
\centering
\fontsize{8}{5}\selectfont
\setlength\tabcolsep{3pt}
\begin{tabular}{lccccccc}
\toprule
\multirow{2}{*}{Dataset} & \multirow{2}{*}{\# subj.}   & \multirow{2}{*}{Manufacturer} & \multirow{2}{*}{Sequence} & Presence of        & Regions with       & \multirow{2}{*}{Resolution (mm)}  & \multirow{2}{*}{Dimension} \\
& & & & abnormalities & expert annotations & & \\
\midrule
 &  &  &  & & Spleen, kidneys, & & \\
&  &  &  &  & gallbladder, esophagus,  &  &  \\
AMOS \cite{Ji2022} & 60 & Philips & MRI (sequences & Abdominal cancer & liver, stomach, aorta, & [0.69×0.69×0.82] & [192×60×64] \\
&  &  & not provided) & and abnormalities & inferior vena cava, pancreas,  & [1.95×3.00×3.00]  & [576×468×512] \\
&  &  &  &  & adrenal glands, duodenum, & & \\ 
&  &  &  &  & bladder, prostate/uterus & & \\ 
\midrule 
 & & & T1 dual in-phase & & Liver, spleen, &  &\\
CHAOS \cite{Kavur2019, Kavur2020, Kavur2021} & 20 & Philips& T1 dual out-phase & Healthy & right kidney, & [1.36×1.36×5.49] & [256×256×26] \\
& & & T2 SPIR & & left kidney & [2.03×2.03×9.00]  & [320×320×50]\\
\midrule
& & GE & T1 & & & &  \\
LiverHCCSeg \cite{Gross2023, LiverHccSeg} & 17 & Philips & arterial & Hepatocellular & Liver (two raters) & [0.74×0.74×1.75] & [256×152×19] \\
& & Siemens & phase & carcinoma & & [1.41×1.41×10.8] & [512×512×131] \\
\bottomrule 
\end{tabular}
\label{tab:eval_datasets}
\end{table*}

\subsection{Benchmarked methods and associated training data}

\subsubsection{MRSegmentator \cite{Häntze2024}} is a method based on nnU-Net \cite{Isensee2021}, and is trained on multiple datasets of various sequences and modalities, including T1-weighted (T1), T2-weighted (T2), and CT scans (Table~\ref{tab:eval_methods}). Training volumes are annotated using an iterative process. First, image-to-image translation is used to convert MRI volumes into pseudo-CT scans. Then, these are segmented with TotalSegmentator \cite{Wasserthal2023}, and the resulting label maps are propagated to the original MRI scans. Finally, the segmentations are manually refined by a radiologist using MONAI Label \cite{DiazPinto2024}. After annotating 50 scans, an initial nnU-Net model is trained and iteratively refined as additional data becomes available. \textit{MRSegmentator} can segment 40 regions\footnote{\url{https://github.com/hhaentze/MRSegmentator}}. We use model v1.2.0 published in August 2024 and implemented in Python 3.11.5.

\subsubsection{MRISegmentator-Abdomen \cite{Zhuang2024}} is also based on nnU-Net, and is trained solely on T1 scans. Similarly to \textit{MRSegmentator}, a cross-modality approach is first used to convert the MRI volumes to synthetic CT scans \cite{Zhuang2024A} in order to obtain labels with TotalSegmentator. Once the labels are propagated back to the MRI volumes, an iterative annotation process is used to progressively train an nnU-Net model. In total, \textit{MRISegmentator-Abdomen} can segment 62 regions\footnote{\url{https://github.com/rsummers11/MRISegmentator}}. We use model v1.0.0 (June 2024, Python 3.11.10).

\subsubsection{TotalSegmentator MRI \cite{D'Antonoli2024}} is a nnU-Net-based architecture extended from TotalSegmentator for CT segmentation \cite{Wasserthal2023}. Starting from manual segmentations of 10 volumes, an iterative strategy is used to train a model, similar to \textit{MRSegmentator} and \textit{\mbox{MRISegmentator-Abdomen}}. In total, 59 regions can be segmented by the model\footnote{\url{https://github.com/wasserth/TotalSegmentator}}. We use model v2.2.0 (May 2024, Python 3.10.13).

\subsubsection{ABDSynth} alleviates the need for manual MRI annotations by leveraging CT segmentations that are already publicly available. Specifically, we use the SynthSeg framework \cite{Billot2023} to train a domain-agnostic network for MRI abdominal segmentation (Figure~\ref{fig:overview}). Synthetic data is generated using a GMM conditioned on the input training label maps. Crucially, the GMM parameters are randomly sampled from uniform distributions of very wide ranges. Moreover, to further increase the diversity of the synthetic data, we apply aggressive augmentations including: affine and non-linear spatial transforms, bias field corruption, contrast augmentation, noise injection, and modeling of various voxel resolutions. Presenting the downstream segmentation network with such data forces it to learn features that are robust against these variations, such that it can segment test scans of any domain without retraining.

We train \textit{ABDSynth} using 128 segmentations from the training set of TotalSegmentator CT \cite{Wasserthal2023}. These label maps are center-cropped/padded to a 300×300×250 size at 1.5mm isotropic resolution. Additional preprocessing details are given in the Appendix. In total, \textit{ABDSynth} segments 33 regions\footnote{\url{https://github.com/deepakri201/AbdoBench}}. We use the same architecture and generation parameters as in \cite{Billot2023}, and train the network for 500,000 iterations with a soft Dice loss \cite{Milletari2016}. Training takes two weeks (Nvidia A100 40GB GPU) using resources provisioned by Jetstream2 \cite{Jetstream1, Jetstream2}.

\begin{table*}[th]
\centering
\fontsize{8}{5}\selectfont
\setlength\tabcolsep{4pt}
\caption{Mean (standard deviations) Dice and HD95 scores obtained by all benchmarked approaches. Methods that are not trained on the tested MRI sequence are in italics. Best-performing methods are in bold, and asterisks denote statistical significance with all other models (5\% level, Bonferroni-corrected Wilcoxon signed-rank test).}
\begin{tabular}{l l c c c c c c c c}
\toprule
Dataset & Region & \multicolumn{2}{c}{MRSegmentator} & \multicolumn{2}{c}{MRISegmentator-Abdomen} & \multicolumn{2}{c}{TotalSegmentator MRI} & \multicolumn{2}{c}{ABDSynth}\\ 
\cmidrule(lr){3-4} \cmidrule(lr){5-6} \cmidrule(lr){7-8} \cmidrule(lr){9-10}
& & Dice & HD95 (mm) & Dice & HD95 & Dice & HD95 & Dice & HD95 \\
\midrule
& Liver                & 0.96 (0.01) & \textbf{2.87}* (1.35) & \textbf{0.97}* (0.02) & 18.86 (40.15) & 0.93 (0.01) & 4.92 (2.35) & \textit{0.95 (0.02)} & \textit{4.50 (4.80)} \\
& Spleen               & 0.94 (0.03) & \textbf{2.04}* (0.98) & \textbf{0.96}* (0.04) & 16.60 (60.13) & 0.90 (0.02) & 2.77 (0.96) & \textit{0.94 (0.05)} & \textit{2.52 (2.94)} \\
& Kidney, left         & \textbf{0.95}* (0.02) & \textbf{1.99}* (0.66) & 0.95 (0.10) & 7.42 (24.74) & 0.91 (0.07) & 2.73 (1.90) & \textit{0.92 (0.06)} & \textit{2.51 (1.33)}  \\
& Kidney, right        & \textbf{0.95}* (0.03) & \textbf{2.25} (1.45) & 0.94 (0.06) & 31.10 (58.37) & 0.92 (0.12) & 3.33 (7.66) & \textit{0.92 (0.13)} & \textit{2.83 (4.98)} \\
AMOS & Pancreas        & 0.80 (0.12) & \textbf{5.20}* (4.95) & \textbf{0.85}* (0.11) & 5.52 (16.54) & 0.75 (0.13) & 6.67 (5.30) & \textit{0.72 (0.17)} & \textit{12.43 (17.88)} \\
& Stomach              & 0.87 (0.10) & \textbf{5.25}* (6.42) & \textbf{0.88}* (0.13) & 11.35 (36.02) & 0.86 (0.10) & 5.86 (7.76) & \textit{0.83 (0.14)} & \textit{8.83 (10.66)} \\
& Gallbladder          & 0.78 (0.18) & 5.78 (6.63) & \textbf{0.82}* (0.18) & 7.40 (20.30) & 0.79 (0.19) & \textbf{5.54}* (7.69)  & \textit{0.70 (0.32)} & \textit{7.65 (8.89)}  \\
& Duodenum            & 0.60 (0.16) & 11.18 (7.46) & \textbf{0.69}* (0.14) & \textbf{9.02}* (5.88) & 0.57 (0.17) & 12.97 (9.97)   & \textit{0.56 (0.22)} & \textit{19.96 (18.32)} \\
& Adrenal gland, left  & 0.53 (0.21) & 6.39 (5.25) & \textbf{0.62}* (0.19) & \textbf{5.39}* (5.64) & 0.51 (0.20) & 6.48 (6.27)  & \textit{0.45 (0.24)} & \textit{9.12 (7.92)} \\
& Adrenal gland, right  & 0.54 (0.14) & 5.88 (3.97) & \textbf{0.61}* (0.13) & \textbf{3.80}* (2.54) & 0.52 (0.14) & 6.04 (3.95) & \textit{0.50 (0.13)} & \textit{6.02 (4.19)} \\
\midrule
& Liver         & \textbf{0.93}* (0.01) & \textbf{2.05}* (0.41) & \textit{0.81 (0.12)} & \textit{40.86 (25.42)} & 0.90 (0.02) & 5.28 (4.45) & \textit{0.91 (0.03)} & \textit{4.86 (5.76)} \\
CHAOS & Spleen  & \textbf{0.89}* (0.02) & \textbf{1.82}* (0.63) & \textit{0.34 (0.23)} & \textit{20.67 (5.64)} & 0.87 (0.02) & 2.01 (0.49) & \textit{0.84 (0.08)} & \textit{3.38 (2.26)} \\
T1 in-phase & Kidney, left & \textbf{0.88}* (0.03) & \textbf{2.29}* (0.49) & \textit{0.63 (0.26)} & \textit{6.83 (9.25)} & 0.79  (0.03) & 3.30 (0.71)  & \textit{0.68 (0.30)} & \textit{8.03 (8.93)} \\
& Kidney, right & \textbf{0.90}* (0.04) & \textbf{1.86}* (0.52) & \textit{0.78 (0.13)} & \textit{18.01 (32.49)} & 0.80 (0.05) & 3.12 (0.50) & \textit{0.68 (0.33)} & \textit{6.45 (6.81)}  \\
\midrule 
& Liver        & \textbf{0.93}* (0.01) & \textbf{2.18}* (0.50) & \textit{0.83 (0.10)} & \textit{30.93 (30.28)} & 0.89  (0.02) & 4.76 (4.22)  & \textit{0.90 (0.03)} & \textit{4.87 (6.09)} \\
CHAOS & Spleen & \textbf{0.88}* (0.03) & \textbf{2.11}* (1.12) & \textit{0.45 (0.18)} & \textit{38.88 (39.16)} & 0.85 (0.03) & 2.29 (0.77)  & \textit{0.76 (0.27)} & \textit{4.25 (5.22)} \\
T1 out-phase & Kidney, left & \textbf{0.87}* (0.04) & \textbf{2.44}* (0.58) & \textit{0.77 (0.06)} & \textit{5.48 (5.82)} & 0.77 (0.04) & 3.52 (0.75)  & \textit{0.74 (0.17)} & \textit{4.18 (2.14)}  \\
& Kidney, right & \textbf{0.88}* (0.04) & \textbf{2.06}* (0.47) & \textit{0.74 (0.18)} & \textit{10.21 (20.21)} & 0.80 (0.03) & 2.95 (0.57) & \textit{0.68 (0.26)} & \textit{4.71 (3.24)}  \\
\midrule
& Liver         & \textbf{0.91} (0.02) & \textbf{2.94} (1.61) & \textit{0.88 (0.12)} & \textit{11.77 (19.06)} & 0.91 (0.03) & 4.59 (6.35)  & \textit{0.90 (0.05)} & \textit{5.11 (6.03)} \\
CHAOS & Spleen       & 0.88 (0.12) & 2.39 (2.50) & \textit{0.87 (0.22)} & \textit{13.56 (28.42)} & 0.86 (0.08) & 5.31 (9.28)  & \textit{\textbf{0.91}* (0.04)} & \textit{\textbf{2.22} (1.67)}  \\
T2 SPIR& Kidney, left  & 0.91 (0.03) & \textbf{2.27}* (0.83) & \textit{\textbf{0.92}* (0.02)} & \textit{5.63 (14.56)} & 0.88 (0.03) & 2.67 (0.86) & \textit{0.83 (0.09)} & \textit{3.34 (0.86)} \\
& Kidney, right & \textbf{0.92}* (0.02) & \textbf{1.88}* (0.77) & \textit{0.91 (0.05)} & \textit{2.69 (2.55)} & 0.90 (0.02) & 2.56 (0.58) & \textit{0.86 (0.14)} & \textit{3.34 (2.25)} \\
\midrule 
LiverHCCSeg &Liver (Rater 1) & \textbf{0.93}* (0.03) & \textbf{4.93}* (3.94) & 0.93 (0.07) & 10.20 (21.90) & 0.91 (0.04) & 6.49 (4.66) & \textit{0.90 (0.08)} & \textit{11.49 (19.42)} \\
\bottomrule
\end{tabular}
\label{tab:scores}
\end{table*}

\subsection{Evaluation datasets}

Our benchmark utilizes three public MRI datasets (Table~\ref{tab:eval_datasets}). These datasets are not used by any of the benchmarked methods for training. Overall, this cohort spans three manufacturers, five MRI sequences, healthy and diseased subjects, and a wide range of resolutions and fields-of-view, all of which enable a comprehensive evaluation for out-of-the-box deployment.

\subsubsection{AMOS \cite{Ji2022}} includes MRI scans of diseased patients with abdominal cancer and other abnormalities acquired at two centers and with eight scanners. Here, we combine the provided training and validation sets (initially used for a Grand Challenge\footnote{\url{https://amos22.grand-challenge.org/}}) into a single test set of 60 volumes. Annotations are provided for 15 abdominal organs (Table~\ref{tab:eval_datasets}). These have been obtained with an iterative process, where a model is first trained on a small set of annotated data to generate pseudo-labels, which are then refined by two radiologists.

\subsubsection{CHAOS \cite{Kavur2019,Kavur2020,Kavur2021}} contains healthy subjects from the Dokuz Eylul University Hospital. It consists of three MRI sequences: T1 dual in-phase, T1 dual out-phase, and T2 SPIR (spectral pre-saturation inversion recovery). The T1 sequences are fat-suppressed, and T2 SPIR is designed to highlight the liver parenchyma\footnote{\url{https://chaos.grand-challenge.org/Data/}}. Consensus ground truth segmentations are provided for the liver, spleen, right kidney, and left kidney using majority voting between three radiologists. 20 volumes are included for each sequence, for a total of 60 volumes. 

\subsubsection{LiverHCCSeg \cite{Gross2023, LiverHccSeg}} includes 17 subjects with hepatocellular carcimona from TCGA-LIHC \cite{Erickson2016} and imaged with a T1 arterial phase sequence. All scans are provided with manual liver segmentations from two independent raters.

\subsection{Evaluation metrics}

We use Dice scores \cite{Dice1945} and the 95th percentile of the Hausdorff distance (HD95) \cite{Huttenlocher1993} as evaluation metrics. Dice quantifies the overlap between two segmented regions in [0,1], where 1 indicates perfect overlap and 0 indicates no overlap. HD95, expressed in millimeters, measures the 95$^\text{th}$ percentile of the surface distance between two segmentations.

%%%%%%%%%%%%%%%%%%%%%%%%%%%%%%%
\section{Results}
%%%%%%%%%%%%%%%%%%%%%%%%%%%%%%%

We organize the benchmark results by datasets (Table~\ref{tab:scores}).

\subsection{AMOS}

\begin{figure*}[t]
\centering
\includegraphics[width=0.9\linewidth]{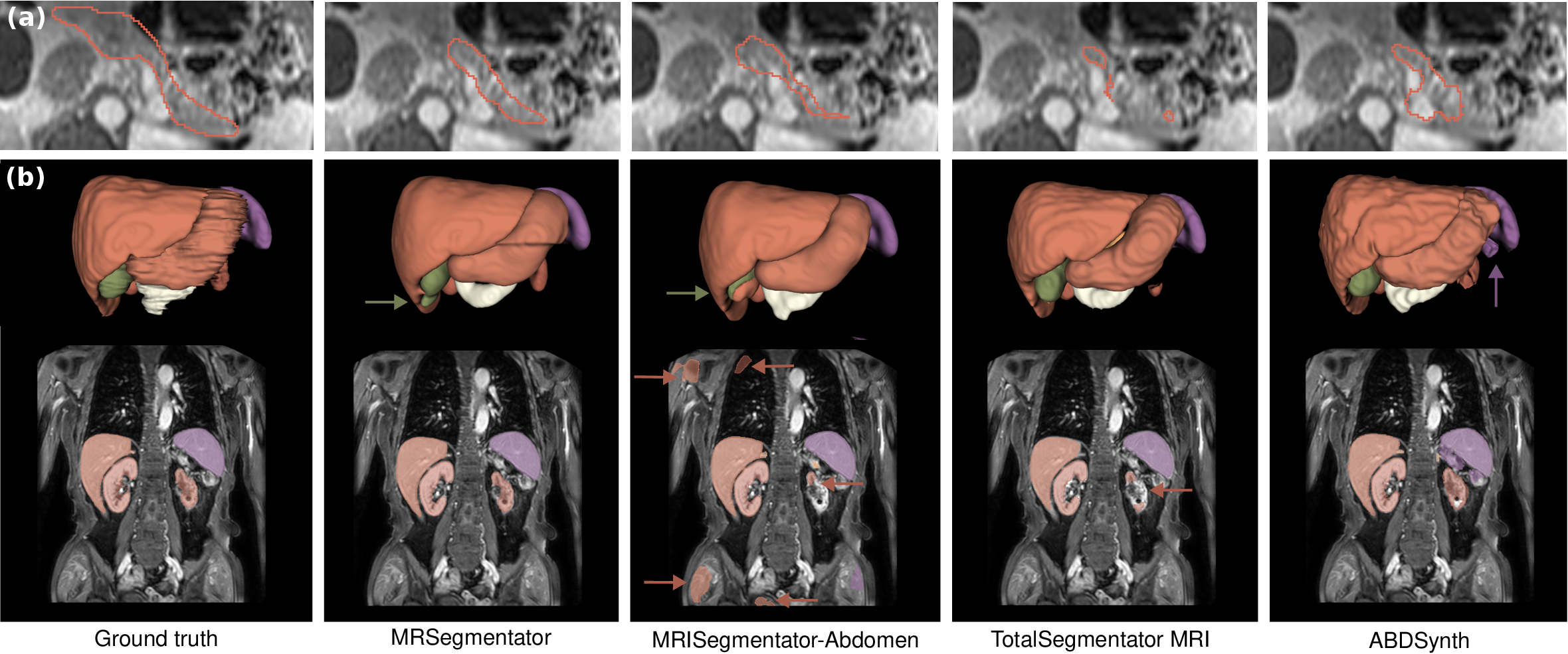} % was 0.85
\caption{Sample segmentations by all methods on AMOS. (a) Pancreas slice, where all methods do not fully segment the region. (b) 3D renderings for another subject. Arrows indicate segmentation errors in the region of corresponding color.}
\label{fig:examples_amos}
\end{figure*}

Among the benchmarked methods, \textit{MRISegmentator-Abdomen} achieves the highest Dice scores for the majority of organs (average gap of 0.035 Dice with \textit{MRSegmentator}), except for the kidneys. However, it consistently exhibits high HD95 values and extreme outliers compared to the other methods. This is likely due to \textit{MRISegmentator-Abdomen} producing implausible segmentations, including predictions of irrelevant regions, as seen in Figure~\ref{fig:examples_amos}. In contrast, \textit{MRSegmentator} and \textit{TotalSegmentator MRI} achieve much more spatially coherent predictions, as indicated by lower HD95 values. \textit{ABDSynth} performs competitively on high-contrast organs such as the liver, spleen, and kidneys, but underperforms on more spatially variable regions, such as the stomach and gallbladder.

\begin{figure*}[t]
\centering
\includegraphics[width=0.9\textwidth]{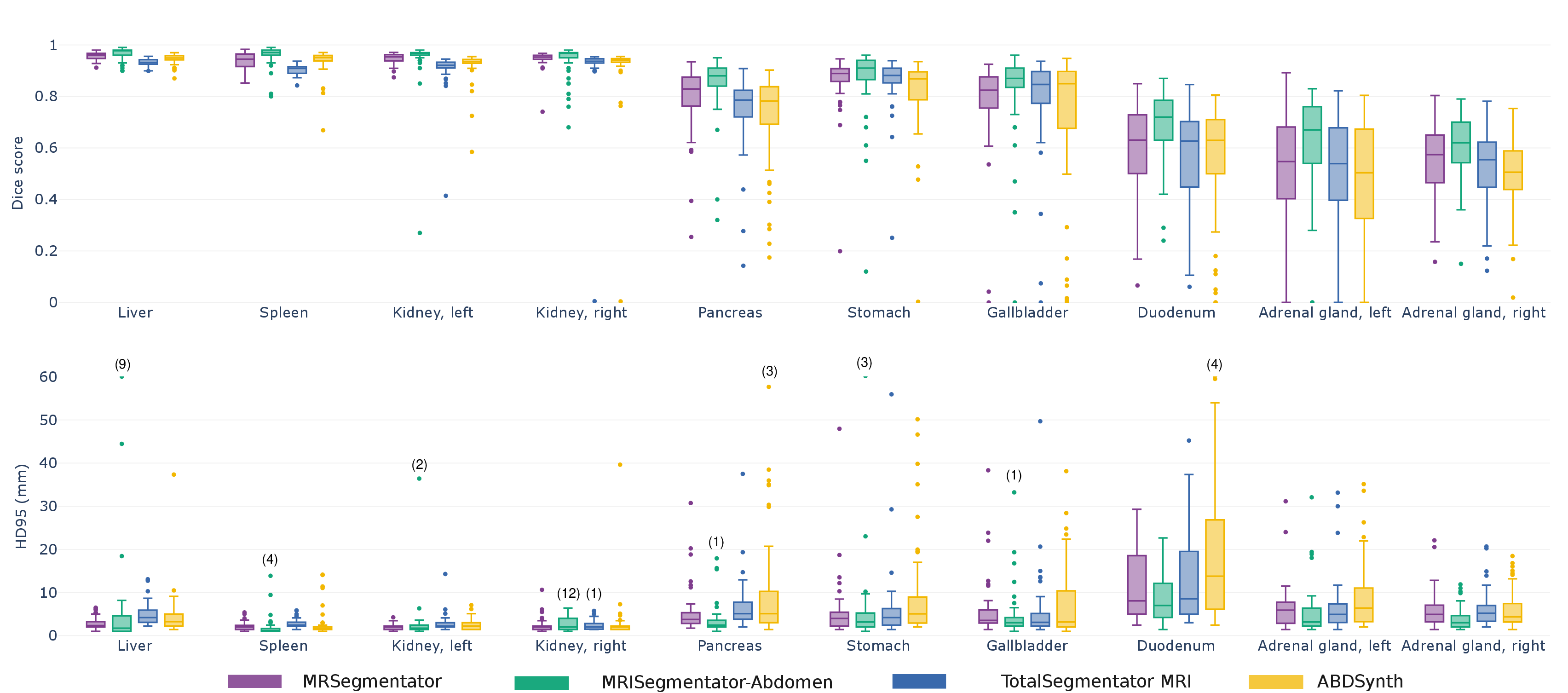} % 0.85
\caption{Dice (top) and HD95 (bottom) boxplots for AMOS results for the four benchmarked methods. We observe similar performances for all methods across the liver, spleen, and kidneys, but highly variable results across the regions that are smaller and/or with more variable morphologies.}
 \label{fig:boxplots_amos}
\end{figure*}

More precisely, Figure~\ref{fig:boxplots_amos} reveals that all four methods yield accurate segmentations of the liver, spleen, and kidneys (Dice higher than 0.9 in all cases), but substantially lower and more variable performances for other regions. This is explained by the morphological variability, small size, and high deformability of regions like the pancreas \cite{Noel2014, Cai2017}. Additional complexity arises for organs such as the duodenum, where peristaltic motion during imaging introduces artifacts that further degrade segmentation performance. Moreover, while smaller Dice scores are indicative of lower performances in the adrenal glands, we highlight that Dice is known to degrade faster in such small regions. This is confirmed by the fact that results are more homogeneous across regions for HD95 than Dice.

Figure~\ref{fig:examples_amos} shows qualitative results for each method, with a focus on the pancreas, where no method achieves anatomically correct segmentation of this region. The 3D renderings reveal over-/under-segmentation (e.g., gallbladder, spleen, and kidneys) patterns across different methods.

\subsection{CHAOS}

\begin{figure*}[t]
\centering
\includegraphics[width=\textwidth]{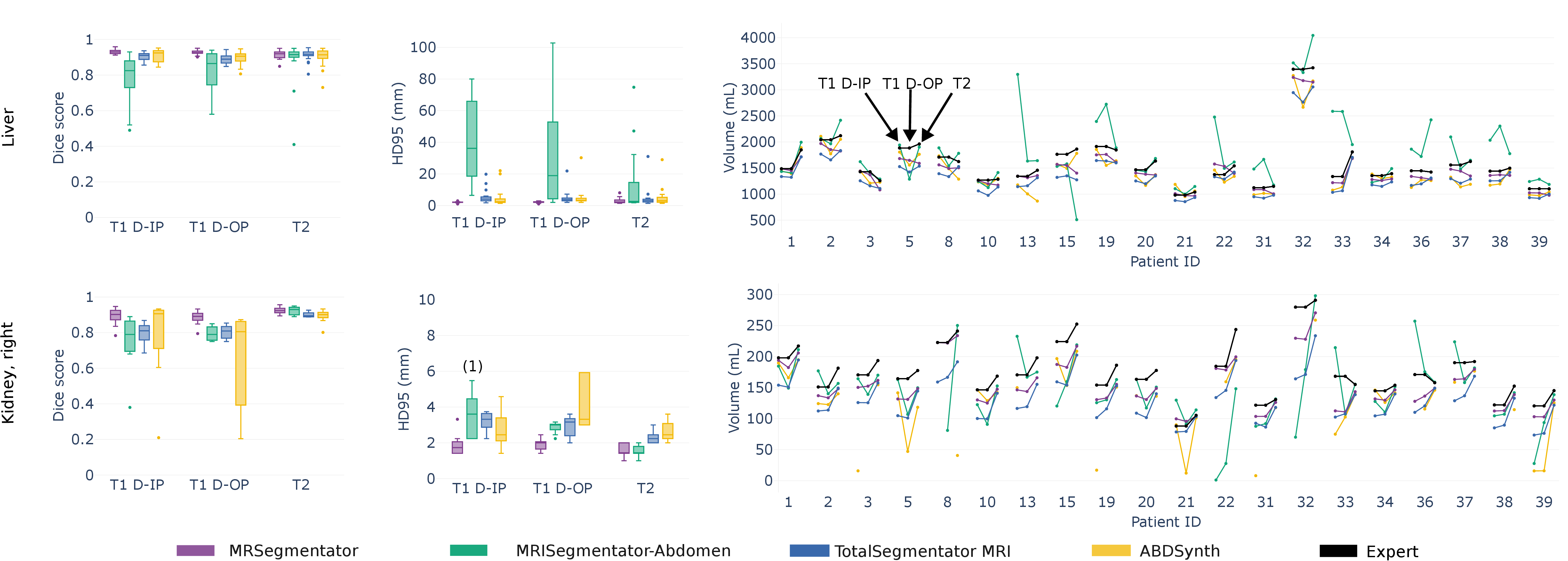} 
\caption{Dice score (left), HD95 (middle), and volume repeatability (right) obtained on CHAOS for two representative regions (liver and right kidney) across different sequences. In the volume repeatability subfigure, the consecutive points represent T1 dual in-phase, T1 dual out-phase, and T2 SPIR, respectively. In general, the liver is more consistently segmented across MRI sequences than the right kidney.}
\label{fig:boxplots_chaos}
\end{figure*}

\begin{figure*}
\centering
\includegraphics[width=0.95\textwidth]{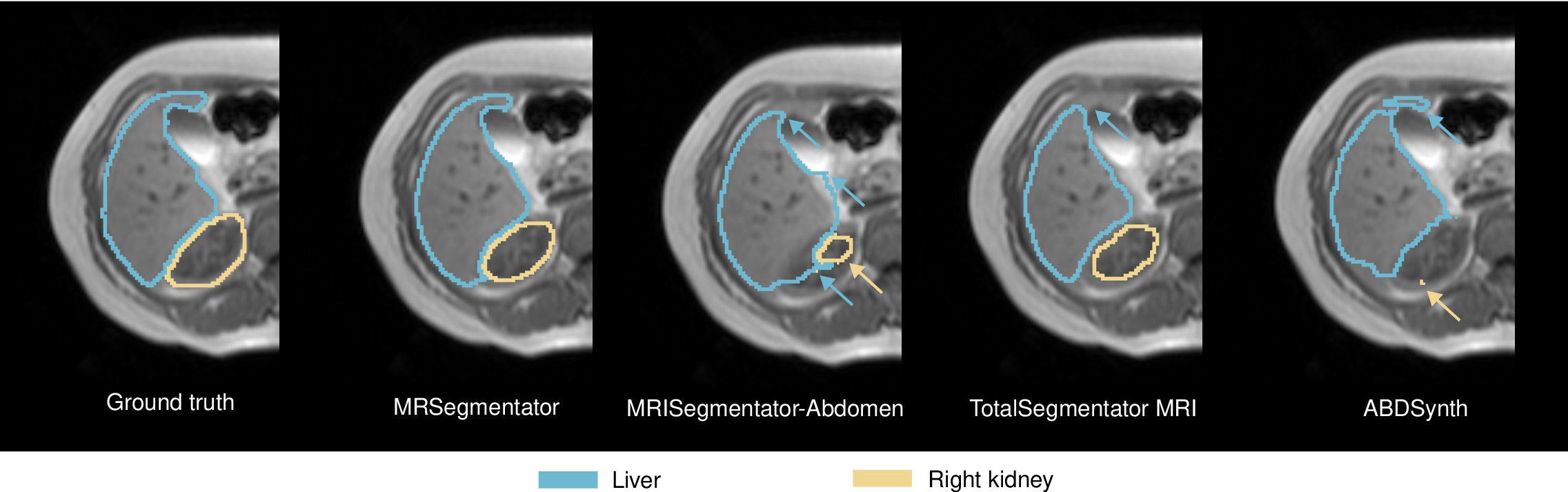} %0.88
\caption{CHAOS subject 39, where blue = liver and yellow = right kidney. Blue and yellow arrows point at major differences between ground truth and automated segmentations for the liver and right kidney, respectively.}
\label{fig:example_chaos}
\end{figure*}

Table~\ref{tab:scores} shows that \textit{MRSegmentator} achieves the highest Dice scores (above 0.87) and lowest HD95 (below 3mm) among all methods and for almost all sequence types and evaluated regions. \textit{TotalSegmentator MRI} and \textit{ABDSynth} also yield fairly high Dice scores and low HD95 values. In contrast, \textit{MRISegmentator-Abdomen}, which is not trained on any of the sequences used in CHAOS, displays much lower Dice scores and higher HD95 values, especially for the T1 scans.

Figure~\ref{fig:boxplots_chaos} illustrates the distributions of the Dice and HD95 metrics, as well as the volume repeatability across the MRI sequences used in CHAOS (T1 dual in-phase/out-phase, and T2 SPIR). In particular, we focus on the liver and right kidney, which are the most and least consistently segmented regions, respectively, among the four available labels in CHAOS.  We also show qualitative segmentation examples obtained by all methods for the liver and right kidney in Figure~\ref{fig:example_chaos}.

For the liver, Dice scores are consistently high across all methods and scan types, with medians above 0.8 in all cases. However, this high overall accuracy is nuanced by the HD95 metric, for which substantial variations in standard deviations indicate local under- and over-segmentations (Figure~\ref{fig:example_chaos}). This effect is particularly visible for \textit{MRISegmentator-Abdomen} (HD95 standard deviations above 19mm for all sequences), where typical segmentation mistakes are illustrated in Figure~\ref{fig:example_chaos}. More precisely, \textit{MRISegmentator-Abdomen}, which has not been trained on any of the CHAOS sequences, yields an average HD95 gap of 16.6mm with the best performing method (\textit{MRSegmentator}) across all CHAOS regions and sequences, which is far worse than the other methods. This issue is also highlighted by the volume analysis (Figure~\ref{fig:boxplots_chaos}), where \textit{MRISegmentator-Abdomen} displays large volumetric intra-subject differences across sequences. In comparison, the other approaches exhibit consistent liver volume estimates across sequences for most subjects, thus highlighting the accuracy of their liver segmentation predictions.

Meanwhile, we observe the opposite trend for the right kidney, where all methods obtain tighter HD95 distributions but more variable Dice scores (Figure~\ref{fig:boxplots_chaos}). Here, the lower Dice scores are due to the substantially smaller size of the kidney, since the Dice metric is more sensitive to segmentation mistakes in smaller regions. Yet, the Dice results also reflect frequent instances of under-segmentation of the right kidney by all methods (Figure~\ref{fig:example_chaos}). In particular, we observe that \textit{ABDSynth} fails to produce segmentations for several subjects, as indicated by missing points in the volume repeatability plot in Figure~\ref{fig:boxplots_chaos}. Nevertheless, these segmentation mistakes remain relatively smaller compared to the liver (all methods produce substantially lower HD95 scores for the right kidney across all sequences), which may be due to the good tissue contrast with the surrounding organs.

\begin{figure*}
\centering
\includegraphics[width=0.95\linewidth]{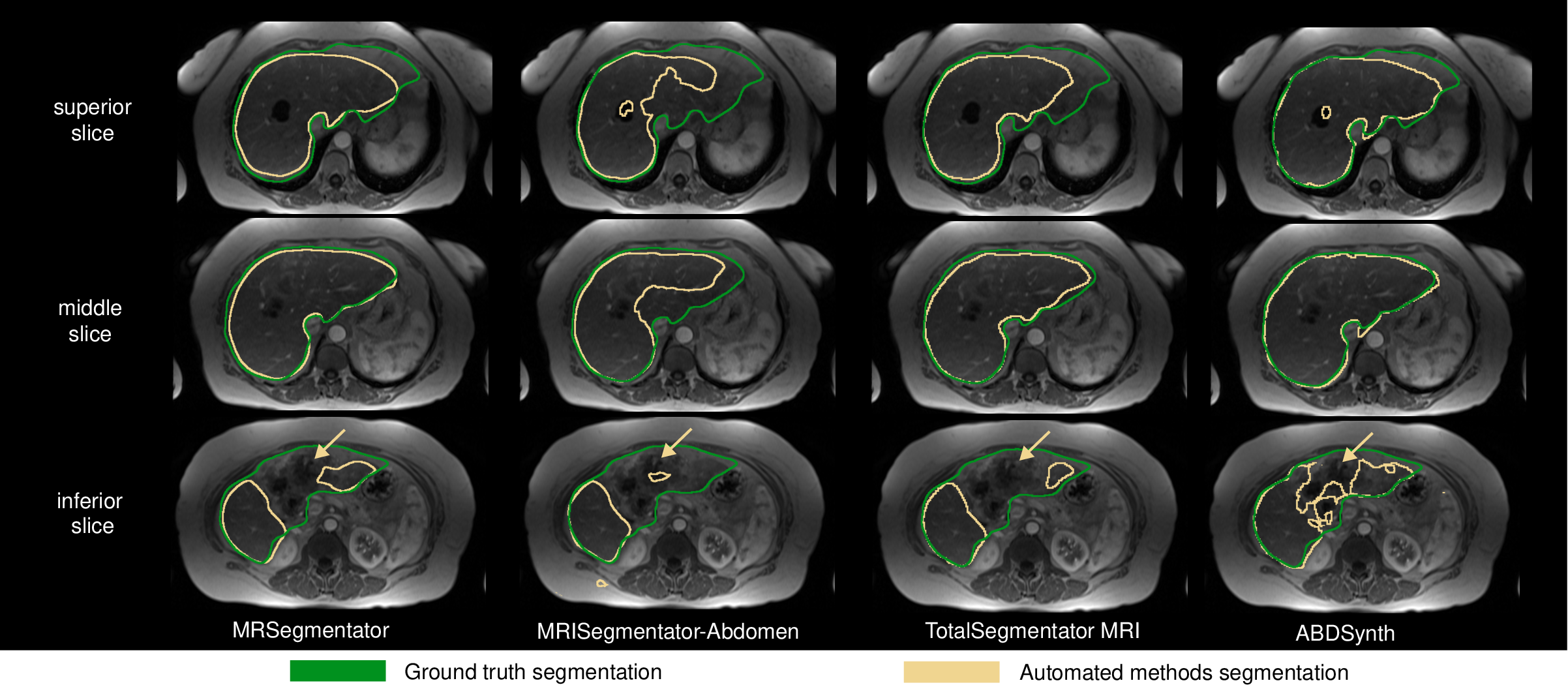} %0.88
\caption{Example of liver segmentations for the superior, middle, and inferior axial slices of a representative LiverHCCSeg subject. For the inferior axial slice, all methods do not segment the liver well, likely due to the presence of a hepatocellular carcinoma, as indicated by the yellow arrows.}
\label{fig:example_liverhccseg}
\end{figure*}

\subsection{LiverHCCSeg}

For the LiverHCCSeg dataset, Table~\ref{tab:scores} presents the performance of automated methods relative to Rater 1. We observe consistently high Dice scores across all methods, ranging from 0.9 (\textit{ABDSynth}) to 0.93 (\textit{MRSegmentator} and \textit{MRISegmentator-Abdomen}). In contrast, the HD95 values exhibit substantial variability, with \textit{ABDSynth} and \textit{MRISegmentator-Abdomen} reporting the highest means at 11.49 mm and 10.20 mm, respectively. Figure~\ref{fig:example_liverhccseg} illustrates this high segmentation variability across methods for a representative subject. It can be seen that while most algorithms perform well in the mid-transverse slice, discrepancies become apparent in the superior slice, where predictions vary widely. In the inferior slice, the presence of a hepatocellular carcinoma appears to degrade segmentation quality across all methods.

Since LiverHCCSeg provides liver annotations from two experts, we now compare the results of all methods against inter-rater reproducibility scores. First, the two experts show a strong overall consistency with a mean Dice score of 0.95 \cite{Gross2023}. Remarkably, all methods yield results that are relatively close, thus highlighting the quality of the produced segmentations. The inter-rater HD is only 15.7mm \cite{Gross2023}, which is worse than any automated method. Beyond further emphasizing the good performance of the benchmarked methods, this result highlights the inter-rater reproducibility issues in annotating regions, especially for diseased tissues such as hepatocellular carcinoma in this example.

\subsection{Computational requirements and inference time}

We now compare all methods in terms of inference time (computed on the same A100 Nvidia GPU as before) and model size (Table~\ref{tab:inference_times}). Inference time differences among the nnU-Net-based models are explained by their preprocessing strategies, and especially by the size of the patches used for sliding-window inference: \textit{MRSegmentator} ($96\!\times\!128\!\times\!160$ patches), \textit{MRISegmentator-Abdomen} ($48\!\times\!160\!\times\!192$), and \textit{TotalSegmentator MRI} ($112\!\times\!128\!\times\!160$). In contrast, \textit{ABDSynth} represents an alternative in time-constrained scenarios, as it is faster (it does not use a patch-based strategy) and is two-thirds smaller in terms of number of parameters. 

\begin{table}[t]
\centering
\caption{Comparison of inference time and model size.}
\fontsize{8}{5}\selectfont
\begin{tabular}{l c c}
\toprule
Benchmarked method & Inference time (s) & Trainable params. \\ 
\midrule
MRSegmentator            & 57.95 $\pm$ 53.15 & 31M \\
MRISegmentator-Abdomen   & 98.19 $\pm$ 69.83 & 31M \\
TotalSegmentator MRI     & 39.60 $\pm$ 10.34 & 31M \\
ABDSynth                 & 21.17 $\pm$ 19.30 & 13M \\
\bottomrule 
\end{tabular}
\label{tab:inference_times}
\end{table}

%%%%%%%%%%%%%%%%%%%%%%%%%%%%%%%
\section{Discussion}
%%%%%%%%%%%%%%%%%%%%%%%%%%%%%%%

In this paper, we present a thorough benchmarking of the state-of-the-art methods in MRI abdominal segmentation: \textit{MRSegmentator}, \textit{\mbox{MRISegmentator-Abdomen}}, and \textit{TotalSegmentator MRI}. Since these methods are trained on MRI segmentations obtained by a labor-intensive iterative process involving several rounds of corrections, we also test another method \textit{ABDSynth} (extending the SynthSeg framework) that only requires widely available CT segmentations to be trained. We perform benchmarking on a collection of three publicly available datasets, AMOS, CHAOS, and LiverHCCSeg, which cover three manufacturers, five different MRI sequences, different subject conditions (healthy and diseased patients), as well as a large range of resolutions and fields-of-view.

\subsection{Robustness of the methods}

\subsubsection{Effect of sequence type}

In order to analyze robustness to different sequences, we focus on the results obtained on CHAOS, which is the only evaluation dataset with multiple sequences for all subjects. Table~\ref{tab:scores} shows that \textit{MRSegmentator} has the highest performance for all sequences. This can be explained by the fact that \textit{MRSegmentator} is trained on the most diverse dataset with multiple T1 and T2 sequences (Table~\ref{tab:eval_datasets}). In comparison, \textit{TotalSegmentator MRI} yields slightly lower performances, which may be due to its less abundant training data (1561 fewer scans than \textit{MRSegmentator}). Regarding \textit{MRISegmentator-Abdomen}, it produces lower quality segmentations on the CHAOS and LiverHCCSeg datasets, which may be due to its relatively high training resolutions compared to the CHAOS resolutions (mean in-slice resolution of 1.28mm vs. 1.62mm, and mean slice spacing of 3.1mm vs. 5mm). Finally, despite \textit{ABDSynth} having never seen real images during training, it can accurately segment them during testing. However, \textit{ABDSynth} fails in some cases since it does not have access to real intensity distributions during training, an issue known as the reality gap \cite{jakobi1995noise}. 

To further study robustness across sequences, we perform statistical tests between the Dice scores obtained by each method on each region of CHAOS by using the Friedman chi-square test with a significance level of 0.05. Almost all of the regions show statistically significant results, except in three cases (\textit{ABDSynth} for liver and spleen, \textit{MRSegmentator} for spleen, and \textit{TotalSegmentator MRI} for spleen), thus emphasizing the remaining performance differences across sequences. 

\subsubsection{Presence of pathologies}

While the CHAOS dataset comprises only healthy subjects, the AMOS and LiverHCCSeg cohorts contain patients with cancer and other abnormalities. Overall, even though the benchmarked methods are trained on datasets containing various abnormalities (e.g., tumors, cysts, etc.), we observe that the evaluated methods may lack robustness to diverse pathological conditions. Importantly, we note that the public datasets used here for benchmarking do not provide subject-specific medical information. Therefore, it is challenging to discern whether a model’s lower performance stems from subject-specific abnormalities or from the inherent difficulty of segmenting certain anatomical regions. This ambiguity is illustrated in Figure~\ref{fig:examples_amos}, which shows consistently poor pancreas segmentation across all methods for a patient with liver pathology, which is likely exacerbated by the pancreas being inherently difficult to segment. Interestingly, for the AMOS dataset, high segmentation performance is observed across all methods for the primary abdominal organs, liver, spleen, and kidneys, as shown in Table~\ref{tab:scores}, despite the presence of pathologies. In contrast, the LiverHCCSeg dataset reveals more variable segmentation quality, likely due to the presence of hepatocellular carcinoma. As illustrated in Figure~\ref{fig:example_liverhccseg}, although all methods achieve high Dice scores (Table~\ref{tab:scores}), segmentation accuracy tends to degrade in inferior slices affected by this pathology, while segmentations of mid-axial slices remain comparatively consistent.

\subsection{Inconsistencies in segmentation conventions}

\begin{figure}
\centering
\includegraphics[width=0.25\textwidth]{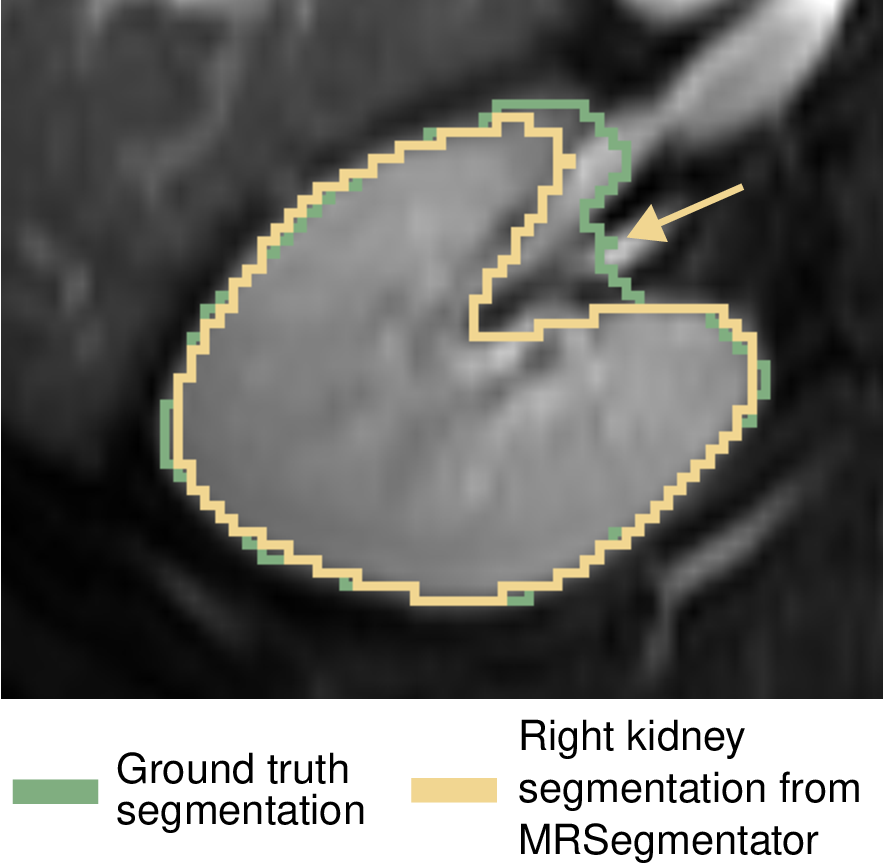} 
\caption{Right kidney segmentations for CHAOS subject 1. The yellow arrow points to major differences between ground truth and automated segmentations, where the ground truth includes the renal pelvis.}
\label{fig:example_chaos_gt_issue}
\end{figure}

Figure~\ref{fig:example_chaos_gt_issue} illustrates differences in anatomical conventions between the CHAOS ground truth annotations and those used in the training data of the evaluated methods. Here, the expert reference includes the renal pelvis as part of the kidney segmentation, whereas the automated methods exclude this region (yellow arrow). These discrepancies in the definition of anatomical boundaries contribute to the lower volume estimates observed in the volume repeatability analysis (Figure~\ref{fig:boxplots_chaos}). This example shows a limitation of our study, where the evaluated methods and benchmarking datasets might use different conventions for some of the regions. More generally, this highlights the importance of identifying semantic inconsistencies in annotation protocols when deploying models across heterogeneous datasets.
% (for instance, the expert annotation of the kidney in CHAOS includes the renal pelvis, while none of the methods were trained on data that included this region).

\subsection{Key difference between benchmarked methods}

Three of the benchmarked methods, \textit{MRSegmentator}, \textit{\mbox{MRISegmentator-Abdomen}}, and \textit{TotalSegmentator MRI} require expert involvement during training, where a clinician or radiologist guides the model through an iterative labels refinement process. In addition to this labor-intensive training paradigm, these methods rely on large datasets: \textit{MRSegmentator} was trained on 2,649 MRI and CT volumes, \textit{\mbox{MRISegmentator-Abdomen}} on 780 MRI volumes, and \textit{TotalSegmentator MRI} on 1,088 MRI and CT volumes. CT data was utilized by \textit{MRSegmentator} and \textit{TotalSegmentator MRI} to improve robustness and cross-modality segmentation capabilities. Furthermore, because these models are trained on specific sequences, their performances degrade when applied to unseen sequences. This limitation is particularly evident in the performance of \textit{\mbox{MRISegmentator-Abdomen}} on the CHAOS dataset (Table~\ref{tab:scores}), where the model shows poorer results across all three scan types, due to a lack of training on those sequences. Consequently, adapting these methods to segment a new MRI sequence would require additional retraining or fine-tuning.

In contrast, \textit{ABDSynth} requires only a single set of annotated CTs, which are widely available, for training. This represents a significant advantage given the relative scarcity of large, annotated MRI datasets compared to CT. By leveraging annotated CT data for MRI segmentation, \textit{ABDSynth} substantially reduces the burden of manual labeling in the MRI imaging space. Furthermore, the method’s synthetic data generation approach enables adaptation to new MRI sequence types without requiring additional expert-annotated MRI datasets and retraining. 

However, as shown in Table~\ref{tab:scores}, \textit{ABDSynth} generally yields lower segmentation performance compared to the other benchmarked methods. Moreover, there are multiple instances where \textit{ABDSynth} produces very poor segmentation outputs, most notably for the T1 out-phase scans in the CHAOS dataset, as reflected by the missing volumes in the volume repeatability plot in Figure~\ref{fig:boxplots_chaos}. These observations highlight an inherent trade-off among the benchmarked methods, balancing \textit{(i)}~training and annotation effort, \textit{(ii)}~the diversity of sequences used for training, and \textit{(iii)}~overall segmentation performance.

%%%%%%%%%%%%%%%%%%%%%%%%%%%%%%%
\section{Conclusion}
%%%%%%%%%%%%%%%%%%%%%%%%%%%%%%%

We presented a benchmarking study of abdominal MRI segmentation methods, including three state-of-the-art models trained on real data and one method trained on synthetically generated data. The models are evaluated on publicly available datasets from multiple grand challenges, as well as a multi-rater liver segmentation dataset. Among the evaluated methods, \textit{\mbox{MRISegmentator-Abdomen}} achieved high Dice scores on the AMOS dataset but exhibited high HD95 values, indicating many outlier segments. Moreover, its performance on the CHAOS dataset was notably lower, likely due to the presence of MRI sequences not included in its training set. In contrast, \textit{MRSegmentator} demonstrated consistent performance across all datasets, with moderately high Dice scores and lower variability, suggesting greater robustness, which is potentially attributed to its more diverse training set. 

Our benchmarking approach suffers from several limitations, which we plan to address in future work. First, while this study focused on a core set of representative automated methods, we did not evaluate all other available tools, and notably TotalVibeSegmentator \cite{Graf2024}, which is specifically designed for segmenting volumetric interpolated breath-hold examination (VIBE) sequences. Inclusion of such specialized models may be considered in future evaluations targeting sequence-specific performance. Secondly, a pathology-specific performance analysis was not feasible due to dataset constraints, as the AMOS dataset lacks pathology labels, CHAOS includes only healthy subjects, and although LiverHCCSeg contains patients who all have hepatocellular carcinoma, it is limited by the small number of subjects in the dataset (17). Future benchmarking efforts would benefit from larger, more diverse datasets with well-annotated pathological labels to enable an evaluation of segmentation performance as a function of different pathologies.

Overall, by releasing our evaluation code as well as the diverse cohort of testing MRI scans, the proposed benchmark represents a first step towards precise and thorough benchmarking of current and future methods for MRI multi-organ abdominal segmentation, a rapidly evolving and promising field for clinical practice.

%%%%%%%%%%%%%%%%%%%%%%%%%%%%%%%
\appendices
\section*{Appendix}
%%%%%%%%%%%%%%%%%%%%%%%%%%%%%%%

\subsection*{Preprocessing the training data of ABDSynth}

In SynthSeg \cite{Billot2023}, synthetic data is generated using a GMM conditioned on training label maps, where each anatomical label is associated with a single Gaussian distribution. While effective, representing the intensities of a given label by a single Gaussian can be insufficient for labels that include heterogeneous substructures with distinct intensity profiles. For example, regions like the renal cortex and medulla in the kidneys, or hepatic vasculature, contain fine-grained differences that are not well captured by a single Gaussian. To address this, we refine the label maps used for synthetic data generation to introduce finer anatomical detail. We adopt a similar strategy to Billot et al. when they extended SynthSeg to cardiac segmentation \cite{Billot2023}. Using the original TotalSegmentator CT scans, we subdivide each label into subregions by clustering the corresponding intensities using expectation-maximization (EM) \cite{Dempster1977}. In order to capture different levels of granularity, we randomly sample the number of clusters from $\{1, 2, 3\}$ for each foreground label. We also apply the same strategy to the background class, but we sample the number of clusters in $\{3, 4, 5, 6, 7\}$ to account for the greater variability of the underlying tissues (Algorithm~\ref{Algorithm1}). Clustering is performed dynamically during synthetic volume generation, with the number of clusters selected at runtime. 

After subdividing labels into substructures, we also simulate different scanning poses that are more specific to MRI acquisitions, and especially poses where only the trunk of the subject is acquired. This is achieved by removing the arms of the subject with a 0.5 probability, where the arm regions have been defined using the 3D Slicer Sandbox extension\footnote{\url{https://github.com/PerkLab/SlicerSandbox\#remove-ct-table}}.

\begin{algorithm}[t]
\caption{Background and foreground labels clustering used for synthetic image generation}
\label{Algorithm1}
\begin{algorithmic}
\For{$K_{BG} \in \texttt{Rand}\{3,4,5,6,7\}$}
    \State $P(x)= \sum_{k=1}^{K_{BG}} \mathcal{N}(x \mid \mu_{k},\sigma^2_k) \pi_k$
    \State Optimize $\theta_{BG}  = \{ \mu_k, \sigma^2_k, \pi_k \mid k=1,\dots,K_{BG} \}$ using Expectation-Maximization (EM) algorithm
\EndFor
\For{label in $[1,\dots,N_{seg}]$}
    \For{$K_{FG} \in \texttt{Rand}\{1,2,3\}$}
        \State $P(x_{\text{label}}) = \sum_{k=1}^{K_{FG}} \mathcal{N}(x_{\text{label}} \mid \mu_k, \sigma^2_k) \pi_k$
        \State Optimize $\theta_{FG}  = \{ \mu_k, \sigma^2_k, \pi_k \mid k=1,\dots,K_{FG} \}$ using EM algorithm
    \EndFor
\EndFor
\State $N_{\text{Seg}}$ refers to the total number of labels used to annotate a specific CT scan.
\State $\pi_k$ refers to the weight of the \textit{k-th} Gaussian component.
\end{algorithmic}
\end{algorithm}

\bibliographystyle{IEEEtran}
\bibliography{references}

\end{document}